\documentclass[pre,twocolumn,superscriptaddress]{revtex4-1}

\usepackage{epsfig}
\usepackage{latexsym}
\usepackage{amsmath}
\usepackage{amssymb}
\usepackage{amsfonts}
\usepackage{graphicx}
\usepackage{wrapfig}
\usepackage[left]{lineno}
\usepackage{url}

\usepackage[normalem]{ulem}
\usepackage{color}

\definecolor{red}{rgb}{1,0,0}
\definecolor{green}{rgb}{0,1,0}
\definecolor{blue}{rgb}{0,0,1}

\begin{document}

\title{Anonymous leadership and stochastic resonance in collectives of self-propelled robots}

\author{M. Dizenhaus}
\affiliation{Instituto Tecnol\'ogico de Buenos Aires (ITBA), Iguazú 341, 1437 C. A. de Buenos Aires, Argentina}%

\author{F. De Simone}
\affiliation{Instituto Tecnol\'ogico de Buenos Aires (ITBA), Iguazú 341, 1437 C. A. de Buenos Aires, Argentina}%

\author{G. A. Patterson}
\email{gpatters@itba.edu.ar}
\affiliation{Instituto Tecnol\'ogico de Buenos Aires (ITBA), CONICET, Iguazú 341, 1437 C. A. de Buenos Aires, Argentina}%

\date{\today}

\begin{abstract}
We investigate the influence of an anonymous leader on a collective of self-propelled robots using Kilobot experiments and numerical simulations. A single leader alternated deterministically between clockwise and counterclockwise motion, while the other robots followed a stochastic majority rule. Although the leader does not change global order, it induces correlations with the collective response that peak at intermediate perturbation levels, resembling stochastic resonance. Simulations confirm that this resonance occurs when the leader's reversal period matches the mean residence time of the unperturbed system. Our results contribute to understanding decision-making in active matter and suggesting principles for steering robotic swarms with minimal leadership input.
\end{abstract}

\maketitle

\section{Introduction}

The study of collective behavior in systems of self-propelled particles has attracted sustained interest in recent decades, spanning disciplines from physics and biology to robotics and social science \cite{ramaswamy2010mechanics,fodor2018statistical}. These systems are characterized by simple local interactions among individuals, which can give rise to complex emergent dynamics such as flocking, swarming, and consensus decision-making. A seminal model in this field is the Vicsek model, which demonstrated that local alignment rules can lead to large-scale collective motion and, under noisy perturbation, revealed an order–disorder phase transition \cite{vicsek1995novel,ginelli2016physics}.

Noise in dynamical systems is not merely a disruptive factor but can also have constructive effects. A prominent example is stochastic resonance (SR), where an optimal level of noise enhances the response of a nonlinear system to a weak periodic signal \cite{gammaitoni1998stochastic}. SR has been observed in diverse contexts, including sensory biology \cite{moss2004stochastic}, neural networks \cite{linder2004effects}, opinion formation models \cite{kuperman2002stochastic,tessone2009diversity,gimenez2022contrarian}, and even resistive switching in electronic devices \cite{patterson2013numerical}. In collective behavior, SR has been reported in network-based decision-making models subjected to oscillating mass-media or propaganda inputs \cite{gimenez2022contrarian,gimenez2023contrarian}. In these studies, the external signal acts as a global field that modulates the state of all agents simultaneously, and the system's response is maximized at intermediate noise levels.

Leadership is another key mechanism that can guide collective behavior. In animal groups, a minority of informed individuals can lead the majority to resources or migration routes without explicit signaling or recognition \cite{reebs2000can,couzin2005effective,king2009leaders}. For instance, Reebs \cite{reebs2000can} showed that trained fish can entrain a shoal of naive individuals to a food source at the correct time of day. Similarly, Couzin et al.~\cite{couzin2005effective} demonstrated that only a small proportion of informed individuals is needed to guide large groups accurately, and that leadership can emerge from differences in information or motivation rather than inherent traits. These findings highlight the concept of anonymous leadership, where influence arises from situational advantages rather than recognized status \cite{strandburg2018inferring,gomez2022intermittent,mugica2022scale}. Leadership is therefore not necessarily tied to fixed identities but can result from contextual factors, such as access to unique information or sudden behavioral changes.

In this work, we investigate how an anonymous leader affects the dynamics of a robotic collective using the experimental setup presented in Ref.~\cite{barone2024experimental}, based on a swarm of Kilobots. These commercial robots have become a widely used platform in collective behavior research, enabling experimental studies of collective transport \cite{rubenstein2013collective}, pattern formation and morphogenesis \cite{rubenstein2014programmable,carrillo2019toward}, decision-making \cite{valentini2016collective}. When Kilobots are programmed with stochastic behavioral rules, nontrivial dynamics can emerge, including bistability in their motion under external fields \cite{patterson2022bistability} and order–disorder transitions in collective behavior \cite{barone2024experimental}. Unlike previous works where leaders change randomly or emerge from fluctuations, here we impose a deterministic periodic leader, while the other agents update their states stochastically according to the rule introduced in Ref.~\cite{barone2024experimental}. We show that although the leader does not significantly alter the global order parameter, it induces correlations between its own state and the collective response. Remarkably, the system exhibits features reminiscent of stochastic resonance: the collective response to the leader is maximized at intermediate perturbation levels, consistent with a matching between the leader's reversal timescale and the mean residence times of the leaderless system.

The paper is organized as follows. In Sec.~\ref{sec:exp}, we present experimental results with Kilobot swarms, characterizing the emergent order and its relation to the leader's dynamics. In Sec.~\ref{sec:num}, we introduce a numerical model that reproduces the experimental observations and allows us to explore longer timescales and parameter ranges. Finally, in Sec.~\ref{sec:con}, we summarize our findings and discuss their implications for understanding leadership, stochastic resonance, and decision-making in active matter and robotic collectives.

\section{Experimental results}
\label{sec:exp}

The experiments were carried out with commercial robots known as Kilobots. They have a diameter of $3.3\ \mathrm{cm}$ and a height of $3.4\ \mathrm{cm}$, and are supported by three rigid legs: one at the front and two at the rear. Locomotion is achieved through two vibration motors independently controlled by an onboard microcontroller. When a motor is activated, the robot rotates around the opposite rear leg: activation of the left (right) motor produces a clockwise (counterclockwise) turn.

In addition to self-propulsion, Kilobots can exchange information with nearby robots via short-range infrared signals, with an effective range of about 12 cm. Each robot broadcasts at a rate of two messages per second. Signal transmission and reception are isotropic, ensuring that any neighbor within range can decode the message regardless of relative orientation. The received information can then be used by the microcontroller to update internal variables or modify the robot's motion. 

The dynamics of the robots rely on alternating clockwise and counterclockwise rotations, following the scheme introduced in Ref. \cite{barone2024experimental}. The direction of rotation of robot $i$ is encoded in an internal variable $\sigma_i$, with $\sigma_i = +1$ for counterclockwise and $\sigma_i = -1$ for clockwise motion. Robots update the value of $\sigma_i$ at fixed intervals $T = 1\ \mathrm{s}$ according to the following stochastic rule:
\begin{align}
\nonumber \sigma_i \leftarrow -\sigma_i&\ ,\ \mathrm{with\ probability\ 1\ if}\ \Phi_i < 0\ ,\\
\label{eq:behavior}\sigma_i \leftarrow -\sigma_i&\ ,\ \mathrm{with\ probability\ }p\mathrm{\ if}\ \Phi_i \geq 0\ ,\\
\nonumber \sigma_i \leftarrow \sigma_i&\ ,\ \mathrm{with\ probability\ }1-p\mathrm{\ if}\ \Phi_i \geq 0\ .
\end{align}    
At each update step, robot $i$ computes the quantity
\begin{align}
\Phi_i = \sigma_i \sum_{j=1}^M \sigma_j\ ,
\label{eq:majority}
\end{align}
where the sum runs over $M$ messages randomly selected from those received during the period $T$. In our experiments we set $M = 2$, so that $\Phi_i$ can take the values $-2$, $0$, or $2$. The value of $\Phi_i$ encodes the relative orientation of robot $i$ with respect to the sampled neighbors: $\Phi_i = -2$ indicates that robot $i$ rotates in the opposite direction to both neighbors, while $\Phi_i = 2$ means that all three robots rotate in the same direction. The intermediate case $\Phi_i = 0$ corresponds to robot $i$ sharing its orientation with one neighbor but not the other, i.e., being aligned with the local majority. The update rule can thus be summarized as a majority rule with stochastic perturbations: a robot always adopts the local majority direction if it is in the minority, while robots aligned with the majority may flip with probability $p$. If no messages are received during $T$, the new state $\sigma_i$ is chosen at random.

Each Kilobot is equipped with a programmable LED that can be used to signal its current $\sigma_i$ value. In our experiments, blue was assigned to $\sigma_i = -1$ and red to $\sigma_i = +1$. The trajectories of all robots were recorded using a top-view camera operating at 1 frame per second. Image processing not only allowed us to extract the trajectories but also to determine the $\sigma_i$ value of each robot by detecting the color of its LED. Each experimental run lasted 60 minutes.

\subsection{Collective behavior}

\begin{figure}
\includegraphics{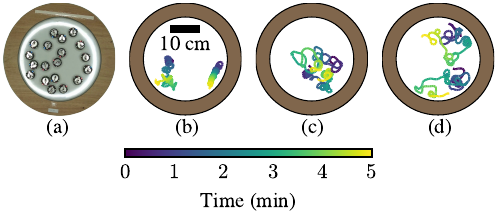}
\caption{\label{fig:experimental} (a) Snapshot of the experimental setup. Nineteen Kilobots were placed inside a circular arena of radius 15 cm. (b)–(d) Trajectories of two robots over a period of 5 minutes for $p = 0.02$, $0.08$, and $0.20$, respectively.}
\end{figure}

We first characterized the collective behavior of a system composed of $N=19$ Kilobots confined within a circular arena of radius 15 cm [Fig.~\ref{fig:experimental}(a)]. The robots were programmed according to Eqs.~\eqref{eq:behavior} and \eqref{eq:majority}, and we studied the emerging dynamics as a function of the control parameter $p$. In accordance with Ref.~\cite{barone2024experimental}, we found that the system undergoes a transition from an ordered state at low $p$ to a disordered state at higher $p$ values. In the ordered regime, the robots exhibit chiral motion, performing localized circular trajectories as shown in Fig.\ref{fig:experimental}(b). As $p$ increases, the trajectories become random, resembling the behavior of active Brownian particles [Figs.~\ref{fig:experimental}(c) and (d)].

The motion of the collective was further characterized through the dynamics of the center of mass (CM). We computed the CM position at each time step, and estimated its velocity $\mathbf{v}_\mathrm{CM}$ and acceleration $\mathbf{a}_\mathrm{CM}$ using finite-difference methods. From these quantities, we calculated the instantaneous angular velocity of the CM,
\begin{align}
\omega_\mathrm{CM}(t) = \frac{\mathbf{v}_\mathrm{CM}(t) \times \mathbf{a}_\mathrm{CM}(t)}{\vert \mathbf{v}_\mathrm{CM}(t)\vert^2}\ ,
\end{align}
where $\omega_\mathrm{CM}>0$ indicates counterclockwise rotation and $\omega_\mathrm{CM}<0$ clockwise rotation. In addition, from the video analysis we obtained the internal states $\sigma_i$ of all robots and used them to define a global order parameter,
\begin{align}
\label{eq:order}
S(t) = \frac{1}{N}\sum_{i=1}^{N}\sigma_i(t)\ ,
\end{align}
where $S(t)=1$ means that all robots rotate counterclockwise, $S(t)=-1$ clockwise, and $S(t)\approx 0$ indicates a balance between both states.

\begin{figure}
\includegraphics{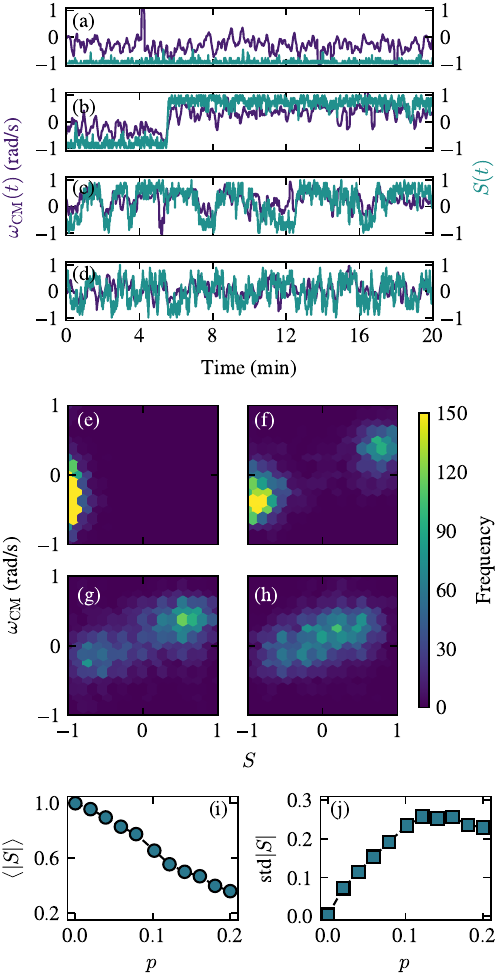}
\caption{\label{fig:timeSL} (a)–(d) Representative 20-minute windows of the temporal evolution of the instantaneous angular velocity of the center of mass and the order parameter of the system for $p = 0.02$, $0.08$, $0.14$, and $0.20$, respectively. (e)–(h) Joint distributions of the angular velocity and the order parameter for the same values of $p$ as in panels (a)–(d), considering the full 60 minutes of each experiment. (i)–(j) Temporal average of the order parameter and its standard deviation as a function of $p$.}
\end{figure}

Figures~\ref{fig:timeSL}(a)–(d) show 20-minute time windows of $\omega_\mathrm{CM}(t)$ for $p=0.02$, $0.08$, $0.14$, and $0.20$, respectively. For the lowest value, $p=0.02$, $\omega_\mathrm{CM}$ remains mostly negative, showing that the CM sustains a persistent clockwise rotation. At $p=0.08$, we observe a spontaneous sign reversal of $\omega_\mathrm{CM}$ around the fifth minute. As $p$ increases further, sign reversals become more frequent (e.g., $p=0.14$), and at the highest value, $p=0.20$, fluctuations are so large that no sustained periods of rotation in a single direction can be identified. In all cases, the figures also display the temporal evolution of the order parameter $S(t)$, which is strongly correlated with the rotation of the CM.

The correlation between $\omega_\mathrm{CM}$ and $S(t)$ is further quantified in Figs.~\ref{fig:timeSL}(e)–(h), which show their joint distributions over the full 60 minutes of each experiment. At $p=0.02$ [Fig.~\ref{fig:timeSL}(e)], the data concentrate in a single region around $S\approx -1$ and negative angular velocities, consistent with a persistent clockwise rotation. At $p=0.08$ and $0.14$ [Figs.~\ref{fig:timeSL}(f)–(g)], two distinct clusters appear, corresponding to spontaneous switches between clockwise and counterclockwise motion. Finally, at $p=0.20$ [Fig.~\ref{fig:timeSL}(h)], the distribution collapses into a central region around $S\approx 0$, characteristic of a disordered state. Despite this, a positive correlation between $\omega_\mathrm{CM}$ and $S(t)$ remains clearly observable, indicating that the global order parameter faithfully captures the collective rotational dynamics of the system.

Finally, for each value of $p$ we computed the time-averaged order parameter, $\langle \vert S(t)\vert \rangle$, together with its standard deviation, $\mathrm{std}\vert S(t)\vert$. Figures~\ref{fig:timeSL}(i) and (j) display the corresponding results. We observe a monotonically decreasing relation between $\langle \vert S(t)\vert \rangle$ and $p$. The behavior of $\mathrm{std}\vert S(t)\vert $ shows a peak, which is characteristic of critical fluctuations and consistent with the second-order phase transition reported in Ref.~\cite{barone2024experimental}.

\subsection{Leadership}

\begin{figure}
\includegraphics{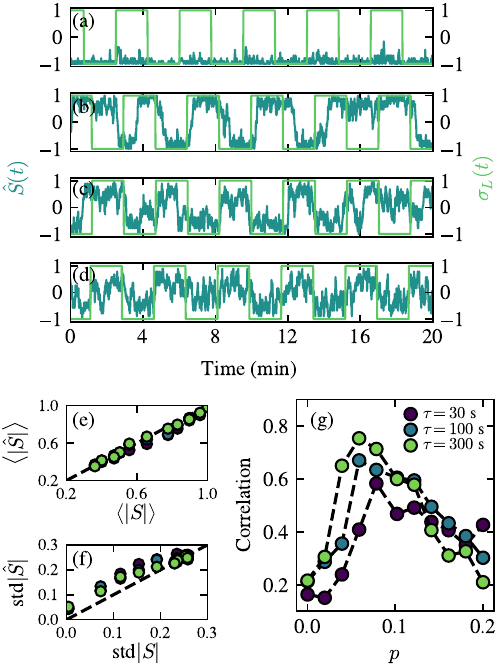}
\caption{\label{fig:timeL} Representative 20-minute windows of the temporal evolution of the collective order parameter and the leader's state for $p = 0.02$, $0.08$, $0.14$, and $0.20$, respectively. Results correspond to a leader that reverses its state with $\tau = 100\ \mathrm{s}$. (e) Time-averaged order parameter: system with leader versus leader-free system. (f) Order parameter standard deviation: system with leader versus leader-free system. (g) Maximum cross-correlation between the system order parameter and the leader's state as a function of $p$. Results are shown for $\tau = 30$, $100$, and $300\ \mathrm{s}$.}
\end{figure}

After characterizing the emergent collective dynamics, we now turn to the main objective of this work: assessing the influence of a leader agent on the system. The leadership we implement is of an anonymous type, meaning that the information transmitted by the leader carries no greater weight than that of any other agent in the group. In addition, the leader is not influenced by its neighbors but instead follows a deterministic behavior, which in our case is chosen to be periodic. Specifically, its internal variable $\sigma_L$ reverses between the values $+1$ and $-1$ at regular intervals of duration $\tau$, as shown in Figs.~\ref{fig:timeL}(a)–(d).

For this part of the study, we considered a collective of 19 robots interacting according to Eqs.~\eqref{eq:behavior} and \eqref{eq:majority}, together with a leader whose reversal period was fixed at $\tau = 30$, $100$, or $300\ \mathrm{s}$. We performed experiments varying the control parameter $p$ that governs the collective dynamics and computed the order parameter $\hat{S}$ using Eq.~\eqref{eq:order}, excluding the state $\sigma_L$ of the leader. Results are shown in Figs.~\ref{fig:timeL}(a)–(d), where 20-minute windows of $\hat{S}(t)$ and $\sigma_L(t)$ are shown for $p = 0.02$, $0.08$, $0.14$, and $0.20$, with $\tau = 100\ \mathrm{s}$. The data reveal that when the collective is strongly organized ($p = 0.02$), the leader has negligible influence. At the other extreme, for large $p$ values, the leader's influence is largely masked by the dominant noise. Strikingly, at intermediate $p$ values ($p = 0.08$ and $0.14$), the order parameter appears to follow, to some extent, the leader's state. This effect becomes more evident when comparing with the leaderless cases in Figs.~\ref{fig:timeSL}(a)–(d): for the same values of $p$, the system exhibits regular transitions between ordered states in the presence of the leader.

We further computed $\langle \vert \hat{S} \vert \rangle$ and $\mathrm{std}\,\vert{\hat{S}}\vert$ and analyzed them against the corresponding quantities obtained in the leaderless case [Figs.~\ref{fig:timeSL}(i)–(j)]. As shown in Fig.~\ref{fig:timeL}(e), the temporal averages distribute along the identity line, indicating that the presence of the leader does not significantly alter the global degree of order for a given $p$. Moreover, the results show that the leader's reversal period $\tau$ has little effect on this quantity. In contrast, the variability is enhanced: Fig.~\ref{fig:timeL}(f) shows that the standard deviation of the order parameter is larger in the presence of the leader, consistent with the higher number of leader-induced transitions between ordered states.

Although the leader does not significantly modify the global order of the system as defined by our chosen order parameter, we do observe a clear correlation between the leader's state and the collective dynamics, suggesting its influence might be captured by an alternative metric. To quantify this relationship, we computed the cross-correlation between $\hat{S}(t)$ and $\sigma_L(t)$ and extracted its maximum value as a function of $p$. The results, shown in Fig.~\ref{fig:timeL}(g), reveal a non-monotonic dependence for all values of $\tau$. The largest amplitude is obtained for $\tau = 300\ \mathrm{s}$. As $\tau$ decreases, the amplitude diminishes and the peak shifts toward higher $p$ values. This behavior is reminiscent of stochastic resonance, where the addition of an optimal level of noise enhances the response of a nonlinear system to a weak external stimulus. The analogy is direct: the leader acts as the weak input signal, the collective as the nonlinear system, and the parameter $p$ provides the stochastic component that maximizes the response. According to the classical definition, synchronization between response and stimulus occurs when the alternation interval of the input signal matches the mean residence time of the unperturbed system. Since this latter quantity depends on the noise intensity, for a fixed reversal period one can maximize the system's response by tuning the level of stochasticity.

Unfortunately, the finite duration of the experiments limits our ability to perform a reliable statistical analysis of the system's residence times. This limitation is illustrated in Fig.~\ref{fig:timeSL}(b), where for $p = 0.08$ the system spontaneously switches its ordered state and then remains stable for more than 15 minutes. For this reason, in order to verify whether the maximization of the response due to the presence of the leader indeed satisfies the matching condition between timescales, we resort to the mathematical model introduced in Ref.~\cite{barone2024experimental}, which allows us to gather a significantly longer temporal evolutions.

\section{Numerical results}
\label{sec:num}

\begin{figure}
\includegraphics{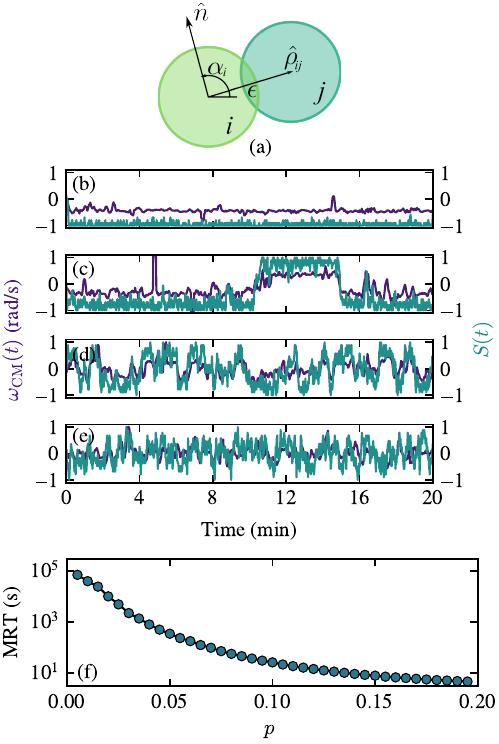}
\caption{\label{fig:simSL} (a) Schematic of a two-particle collision illustrating the relevant quantities. (b)–(e) Simulated 20-minute windows of the temporal evolution of the instantaneous angular velocity of the center of mass and the order parameter of the system for $p = 0.02$, $0.08$, $0.14$, and $0.20$, respectively. (f) Mean residence time as a function of $p$.}
\end{figure}

\begin{figure}
\includegraphics{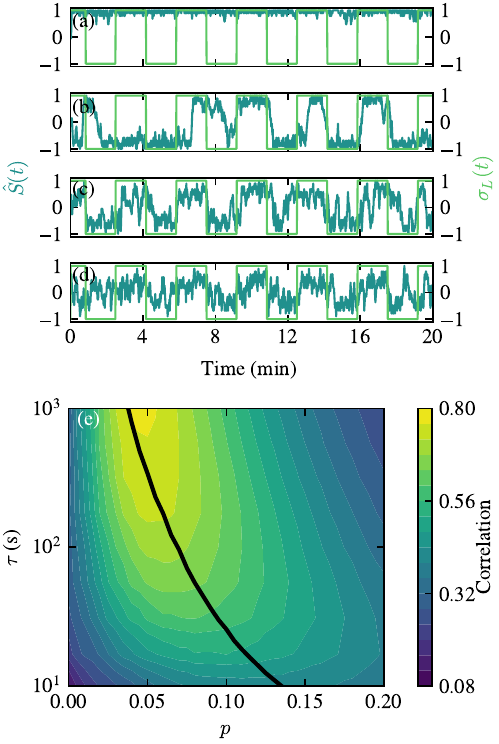}
\caption{\label{fig:simL} Simulated 20-minute windows of the temporal evolution of the collective order parameter and the leader's state for $p = 0.02$, $0.08$, $0.14$, and $0.20$, respectively. Results correspond to a leader that reverses its state with $\tau = 100\ \mathrm{s}$. (e) Maximum cross-correlation between the system order parameter and the leader's state as a function of $p$ and $\tau$. The black line indicates the MRT of the system in the absence of a leader revealing resonance-like behavior consistent with timescale matching.}
\end{figure}

The mathematical model used in this study follows the particle dynamics described in Ref.~\cite{barone2024experimental}. In this model, particles move by rotating around an eccentric axis in one of two possible directions under overdamped dynamics. Each particle, of radius $R_p$, moves at a constant velocity $v_0$ when there is no contact with other particles, and its orientation is defined by $\hat{n}_i=(\cos\alpha_i, \sin\alpha_i)$, as illustrated in Fig.~\ref{fig:simL}(a). The overdamped dynamics of motion and orientation of particle $i$ are given by:
\begin{align}
\frac{d\vec{r}_i}{dt} &= v_0 \hat{n}_i + \sum_j \vec{F}_{ij}\ ,\label{eq:velocity} \\
\frac{d\alpha_i}{dt} &= \sigma_i(t)\frac{|v_i|}{R_p} + \eta_i\ ,\label{eq:angular} 
\end{align}
where $\vec{r}_i$ denotes the position of the particle and $\sum_j \vec{F}_{ij}$ represents the effective interaction force arising from collisions with other particles and/or the boundaries of a possible container. This force has intensity $\kappa$ and is proportional to the overlap length $\epsilon$ between the two contacting objects. Its direction is given by the normal contact vector $\hat{\rho}_{ij} = \frac{\vec{r}_j - \vec{r}_i}{|\vec{r}_j - \vec{r}_i|}$, such that the contact force reads
\begin{eqnarray}\label{eqForce}
    \vec{F}_{ij} = -\kappa \epsilon \hat{\rho}_{ij}\ .
\end{eqnarray}
Figure~\ref{fig:simSL}(a) shows a collision between two particles illustrating the overlap length $\epsilon$ and the contact vector $\hat{\rho}_{ij}$.

Equation~\eqref{eq:angular} governs the evolution of each particle's orientation. As in the experiments, $\sigma_i(t)$ is the internal motion state, taking values $\pm 1$ depending on whether the particle rotates clockwise or counterclockwise. The first term in the equation causes the particle to rotate with angular speed $\vert v_i \vert / R_p$ in the direction indicated by $\sigma_i(t)$. An additional uncorrelated noise term, $\eta_i$, accounts for fluctuations induced by vibration; this is modeled as a Gaussian random variable with zero mean and amplitude $A$. $\sigma_i(t)$ is updated in the same way as in experiments: at discrete intervals $T=1\ \mathrm{s}$, the particle samples the states of its neighbors (within a radius of $12\ \mathrm{cm}$) and decides whether to maintain or invert its current rotation direction, according to the rule defined by Eqs.~\eqref{eq:behavior} and~\eqref{eq:majority}.

The implementation of the leader particle was the same as in experiments: a particle that is not affected by the state of its neighbors and whose internal state $\sigma_L$ alternates deterministically at fixed intervals $\tau$.

We performed simulations lasting $10^5\ \mathrm{s}$ using the Euler–Maruyama algorithm with an integration step $\Delta t = 10^{-2}\ \mathrm{s}$. The model parameters were the same as in Ref.~\cite{barone2024experimental}: $v_0 = 8.25\ \mathrm{mm/s}$, $R_P = 1.65\ \mathrm{cm}$, $\kappa = 50\ \mathrm{1/s}$, and $A=0.05$. The container radius was $15\ \mathrm{cm}$, identical to the experimental setup. Both in the absence and in the presence of a leader, we varied the control parameter $p$ in the range $[0.0, 0.2]$. For the simulations with a leader, we also explored different values of $\tau$ in the range $[10, 1000]\ \mathrm{s}$.

Figures~\ref{fig:simSL}(b)–(e) show the temporal evolution of the instantaneous angular velocity of the system's center of mass [$\omega_\mathrm{CM}(t)$] and the order parameter [$S(t)$] for $p=0.02$, $0.08$, $0.14$, and $0.20$. We observe that the numerical results are in excellent agreement with the experiments [see Fig.~\ref{fig:timeSL}(a)–(d)]. The longer duration of simulations also allows us to compute the mean residence time (MRT) in ordered states. For this, we measured the time intervals during which $S(t)$ maintained the same sign. Figure~\ref{fig:simSL}(f) shows the MRT as a function of $p$. The results reveal a monotonically decreasing relation: at low values of $p$, transitions between states are rare, while at higher values they become increasingly frequent, consistent with the experimental observations.

We then extended the study to include a leader particle. Figures~\ref{fig:simL}(a)–(d) show the temporal evolution of the system's order parameter $\hat{S}(t)$ and the leader's state $\sigma_L(t)$ for a leader that reverses its state with $\tau = 100\ \mathrm{s}$. Once again, the model successfully reproduces the experimental behavior [see Figs.~\ref{fig:timeL}(a)–(d)]. From the cross-correlation between $\hat{S}(t)$ and $\sigma_L(t)$, computed for each value of $p$ and $\tau$ and averaged over 300 realizations, we obtained the results shown in Fig.~\ref{fig:simL}(e). These reveal a non-monotonic dependence between the leader–system correlation and the analyzed parameters. Specifically, the system's response is maximized at intermediate values of $p$, and the maximum correlation increases with $\tau$. Additionally, the correlation peak shifts toward higher values of $p$ as $\tau$ decreases. These results generalize the experimental findings reported in Fig.~\ref{fig:timeL}(g). Finally, Fig.~\ref{fig:simL}(e) also includes the MRT curve as a function of $p$, enabling a direct comparison with the timescale-matching condition that characterizes stochastic resonance. Although the matching criterion is a theoretical approximation \cite{gammaitoni1998stochastic}, the agreement is remarkably good. This amplification of the system's response demonstrates that a leader's influence is optimal when the collective operates in a critical region between order and disorder.

\section{Conclusion}
\label{sec:con}

We studied the collective dynamics of self-propelled robots using Kilobot experiments and numerical simulations. The system is characterized by a control parameter $p$ that introduces perturbations in the interaction mechanism. As $p$ increases, the system transitions from ordered to disordered states, reflected in the collective motion: ordered states produce coherent rotation of the center of mass, while disordered states lead to random movement.

The introduction of a single leader that periodically alternates its state does not alter the collective order level but synchronizes the group's transitions with those of the leader. These correlations exhibit a resonance-like dependence on the control parameter: at intermediate $p$, the system maximizes its response, consistent with a stochastic resonance mechanism, which occurs when the leader's reversal timescale matches the mean residence time of the unperturbed system in ordered states.

Our results extend previous work on stochastic resonance in network and opinion models \cite{kuperman2002stochastic,tessone2009diversity,gimenez2022contrarian}, showing that such effects can emerge from a single embedded agent rather than an external field. They also connect with observations in biological collectives, where anonymous leadership can drive coherent group behavior \cite{mugica2022scale}. Overall, our study provides evidence that weak deterministic inputs can enhance collective responsiveness, offering design principles for robotic swarms based on the interplay between interaction, noise, and leadership.

\section*{ACKNOWLEDGMENTS}
This work was funded by project PICTO-2022-ITBA-00001 (Agencia Nacional de Promoci\'on Cient\'ifica y Tecnol\'ogica, Argentina and Instituto Tecnol\'ogico de Buenos Aires, Argentina)

\bibliography{manusBib}

\end{document}